\documentclass[aps,pra,showpacs,eqsecnum,twocolumn]{revtex4}
\usepackage{graphicx}
\usepackage{bm}
\usepackage{amssymb, amsmath}
\usepackage{color}
\usepackage{subfigure}
\usepackage{hyperref}

\begin{document}

\preprint{}
\title{How general are time-local master equations?}
\author{Daniel Maldonado-Mundo, Patrik \"Ohberg, Brendon W. Lovett and Erika Andersson}
\affiliation{SUPA, Department of Physics, 
Heriot Watt University, 
Edinburgh EH14 4AS, United Kingdom}
\date{\today}
\begin{abstract}
	
Time-local master equations are more generally applicable than is often recognised, but at first sight it would seem that they can only safely be used in time intervals where the time evolution is invertible. Using the Jaynes-Cummings model, we here construct an explicit example where two different Hamiltonians, corresponding to two different non-invertible and non-Markovian time evolutions, will lead to arbitrarily similar time-local master equations. This illustrates how the time-local master equation on its own in this case does not uniquely determine the time evolution. The example is nevertheless artificial in the sense that a rapid change in (at least) one of the Hamiltonians is needed. The change must also occur at a very specific instance in time. If a Hamiltonian is known not to have such very specific behaviour, but is ``physically well-behaved", then one may conjecture that a time-local master equation also determines the time evolution when it is not invertible. 

\end{abstract}


\maketitle

\section{Introduction} 
\label{sec:introduction}

In Nature, quantum systems are never truly isolated; the dynamics of a quantum system  depends, to lesser or greater extent, on its interaction with an environment~\cite{BreuerPetruccione}. For an isolated system, the evolution would be unitary, determined by the system Hamiltonian. If the system is coupled to an environment, and system and environment together undergo unitary evolution, then the result is non-unitary evolution of the reduced density matrix of the system. Tracing over the environment in the Schr\"odinger equation for system and environment gives a  first order differential equation for the reduced density matrix of the system. In this so-called master equation, the environment enters the description only as parameters, which may be time-dependent, but are nevertheless not variables. 

If correlations between the system and the environment are short-lived, then the system has no memory of its previous evolution, other than what is contained in the present state of the system itself. The time evolution then has a semi-group structure and the master equation can be written in standard Lindblad form~\cite{Lindblad,GoriniKossakowskiSudarshan}. Such time evolution is said to be Markovian, has been much investigated and is well understood.

Non-Markovian time evolution occurs, somewhat loosely speaking, when the future evolution of a quantum system is affected by its past history. This may happen if there is feedback from the environment onto the system, which influences the future evolution. However, non-Markovian behaviour is also possible when the environment is unaffected by the system, and there thus can be no feedback in the usual sense~\cite{LoFrancoAnderssonBellomo}. Non-Markovian processes appear in quantum optics \cite{BreuerPetruccione,GardinerZoller}, solid state physics \cite{Coish2004} and quantum information processing \cite{AharonovKitaevPreskill}. Even though non-Markovian behaviour appears in many branches of physics, there is no definition of non-Markovianity that is generally agreed upon. Several measures of non-Markovianity  have been proposed~\cite{AnderssonCresserHall2010,BreuerLainePiilo,WolfEisertCubbitCirac, RivasHuelgaPlenio} and compared~\cite{HaikkaCresser,Chrusacutecinacutesk2011}. 

A master equation in Lindblad form is time-local. This is a useful way of writing master equations since it, in particular, enables the use of quantum trajectory methods~\cite{Breuer2004}. At first sight, it might seem that time-local master equations can only describe Markovian evolution. Appearances deceive, however, and they can in fact also describe non-Markovian behaviour~\cite{BreuerPetruccione}. As long as the time evolution is invertible (and in some other special cases), it can be uniquely described using a time-local master equation~\cite{AnderssonCresserHall2010}. If the time evolution is not invertible, then time-local master equations typically cease to uniquely determine the time evolution. One must then also be careful when using numerical techniques such as quantum trajectory methods.

In this paper we will give an explicit example where two different time evolutions, corresponding to two different Hamiltonians, both lead to the same time-local master equation. The time evolutions are explicitly non-Markovian and are constructed using the Jaynes-Cummings model \cite{JaynesCummings} with time-dependent coupling. The Hamiltonian for one of the time evolutions changes rapidly at the specific time when the time evolution becomes non-invertible. We conjecture that if it can be guaranteed that the Hamiltonian does not have such ``artificial'' features, then the time evolution is in principle uniquely determined by a time-local master equation even in cases when the time evolution is not invertible. This would broaden the applicability of time-local master equations to encompass an even larger class of physical time evolutions.

In Section \ref{sec:master_equations}, we review concepts related to master equations and time evolution, and the Jaynes-Cumming model. We then proceed to construct and investigate the two time evolutions with the same time-local master equation in Section \ref{sec:master_equation_for_a_2_level_atom}. We finish with conclusions.


\section{Master equations and time evolution} 
\label{sec:master_equations}

Master equations describe the evolution of a quantum system $S$ coupled to an environment $E$. Under fairly general conditions, a master equation for the density matrix of the system takes the form~\cite{BreuerPetruccione}
\begin{equation}
	\dot{\hat{\rho}}\left(t\right)=-\frac{i}{\hbar}\left[\hat{\mathcal{H}},\hat{\rho}\left(t\right)\right]+\int_{0}^{t}\mathcal{K}_{u,t}\left[\hat{\rho}\left(u\right)\right]du,
	\label{EqGenFormME}
\end{equation}
where $\hat{\mathcal{H}}$ is the system Hamiltonian, and the memory kernel $\mathcal{K}_{u,t}$ is a linear map describing the effects of the environment on the system.

The Born-Markov approximation amounts to taking the kernel as $\mathcal{K}_{u,t}\left[\hat{\rho}\left(u\right)\right]\approx\mathcal{K}\delta\left(t-u\right)\hat{\rho}\left(u\right)$, and assuming that the coupling between the system and the environment is weak. 
Then there is no memory, and the system dynamics are said to be Markovian. This leads to a master equation in Lindblad form~\cite{Lindblad, GoriniKossakowskiSudarshan},
\begin{eqnarray}
	\label{EqMEqLindbladForm}
		 \dot{\hat{\rho}}\left(t\right) &= & -  \frac{i}{\hbar}\left[\hat{\mathcal{\widetilde{H}}},\hat{\rho}\left(t\right)\right] \\*
	&& -  \sum_{k}\frac{\gamma_{k}}{2}\left(\hat{L}_{k}^{\dagger}\hat{L}_{k}\hat{\rho}\left(t\right)+\hat{\rho}\left(t\right)\hat{L}_{k}^{\dagger}\hat{L}_{k}-2\hat{L}_{k}\hat{\rho}\left(t\right)\hat{L}_{k}^{\dagger}\right),\nonumber 
\end{eqnarray}
where $\gamma_{k}\geq0$ are the decoherence rates, the operators $L_{k}$ describe the different decoherence channels, and the Hamiltonian term may now include effects arising from the environment.

The master equation (\ref{EqMEqLindbladForm})  has a time-local form, that is, the RHS is a linear operator acting on $\hat\rho(t)$ and does not explicitly involve the past history of the system. Nevertheless, time-local master equations can describe also non-Markovian time evolution~\cite{BreuerPetruccione}. A simple motivation for this is given in \cite{AnderssonCresserHall}. One uses the fact that any physical time evolution in quantum mechanics is described by a completely positive map~\cite{BrussLeuchs}
\begin{equation}
	\hat\rho(t) = \phi_{t}[\hat{\rho}(0)]=\sum_{k}\hat{A}_{k}(t)\hat{\rho}(0)\hat{A}_{k}^{\dagger}(t),
	\label{EqKrausDefinition}
\end{equation}
	where $\hat{A}_{k}(t)$ are time dependent Kraus operators which satisfy $\sum_{k}\hat{A}_{k}^{\dagger}(t)\hat{A}(t)=\hat{\mathcal{I}}$, guaranteeing that the map is trace-preserving. The solution of a master equation should be a valid physical time evolution, and hence it can be written as a CP map \footnote{Sometimes, because of the approximations made when deriving a particular master equation, its solution may show unphysical features, and then also cannot be described as a CP map. For example, for certain initial states, the density matrix may acquire negative eigenvalues for short initial times~\cite{Ambegaokar, GardinerMunro, BarnettStenholm}. We can however rewrite (\ref{EqGenFormME}) as given in \eqref{EqCPMapVSLinearMap} as soon as $\hat\rho(t)$ depends on the initial state $\hat{\rho}(0)$ via a map $\phi_t$ that is invertible in the time interval studied; $\phi_t$ does not need to be completely positive. }.
We then have
\begin{eqnarray}
	\dot{\hat{\rho}}(t) & = & \int_{0}^{t}\mathcal{K}_{s,t}[\hat{\rho}(s)]\,ds \nonumber \\*
	& = & \int_{0}^{t}\left(\mathcal{K}_{s,t}\circ\phi_{s}\right)[\hat{\rho}(0)]\,ds \nonumber \\*
	& = &  \int_{0}^{t}\left(\mathcal{K}_{s,t}\circ\phi_{s}\circ\phi_{t}^{-1}\right)[\hat{\rho}(t)]\,ds,
		\label{EqCPMapVSLinearMap}
\end{eqnarray}
where we have omitted the trivially time-local Hamiltonian term. From this we obtain the time-local master equation
\begin{equation}
	\dot{\hat{\rho}}(t)=\Lambda_{t}\left[\hat{\rho}(t)\right],
	\label{EqTimeLocalDef}
\end{equation}
where the map $\Lambda_{t}$ can be identified with $\int_{0}^{t}\mathcal{K}_{s,t}\circ\phi_{s}\circ\phi_{t}^{-1}ds$.
Therefore, a time-local formulation is in principle possible as soon as the time evolution is invertible in the time interval considered. This includes non-Markovian cases. If the time evolution is not invertible in the time interval considered, i.e., if at least two different states evolve to the same state at some time $t$, then the the inverse map $\phi^{-1}_t$ does not exist, and the above argument breaks down \footnote{In this case, it would seem that one cannot describe the evolution by a time-local master equation. In special cases, loosely speaking when the time evolution never re-enters a part of the Hilbert space that it previously exited completely, a time-local master equation may still be possible also for non-invertible time evolution~\cite{AnderssonCresserHall}. In this paper, we consider the case of non-invertible time evolution more generally.}. 

Non-Markovian systems thus admit a description using Lindblad-like master equations of the form (\ref{EqMEqLindbladForm}), but with time-dependent decay rates $\gamma_{k}(t)$ and decoherence channels $\hat{L}_{k}(t)$~\cite{Breuer2004,AnderssonCresserHall2010}, which may also depend on the initial state of the environment and on the initial state of the system itself. In what is sometimes called the time-dependent Markovian case, all $\gamma_{k}(t)\geq0$ for all times. For the truly non-Markovian case, one or more of the $\gamma_{k}(t)$ takes negative values for certain time intervals. Negative decay rates may, for example, correspond to processes where the system is reversing to its initial state.  Several measures have been proposed to characterise non-Markovianity \cite{BreuerLainePiilo,AnderssonCresserHall2010,WolfEisertCubbitCirac, RivasHuelgaPlenio}. There are slight differences between these. For example, evolution which according to the trace-distance measure~\cite{BreuerLainePiilo} is ``Markovian'' may be ``non-Markovian'' in the sense that negative decay rates appear in the master equation~\cite{HaikkaCresser}. This is because the effect of the ``non-Markovian'' decay channels, as far as trace distances are concerned, may be cancelled out by other ``Markovian'' decay channels with positive decay rates \footnote{When determining whether or not there are negative decay rates in a master equation of the form (\ref{EqMEqLindbladForm}), it is important to use its diagonal (canonical) form. Otherwise, terms which appear to have negative decay rates may be cancelled out, in a non-trivial way, by other terms with positive decay rates~\cite{AnderssonCresserHall2010}.}.

\section{Master equation for a two-level system}
\label{sec:master_equation_for_a_2_level_atom}

We will now construct a non-invertible time evolution and the corresponding time-local master equation, with the aim of investigating the extent of the validity of this master equation. Let us suppose that a two-level quantum system with excited state $|e\rangle$ and ground state $|g\rangle$ evolves according to a CP map $\hat{\rho}(t)=\hat{A}_{1}(t)\hat{\rho}(0)\hat{A}^{\dagger}_{1}(t)+\hat{A}_{2}(t)\hat{\rho}(0)\hat{A}^{\dagger}_{2}(t)$, where the Kraus operators are~\cite{AnderssonCresserHall}
\begin{eqnarray}
	\hat{A}_{1}(t) = & |g\rangle\langle g|+f(t)|e\rangle\langle e|, \nonumber\\ 
	\hat{A}_{2}(t) = & \sqrt{1-|f(t)|^{2}}|g\rangle\langle e|.
	\label{EqKrausOperators}
\end{eqnarray}
Continuity of $\hat\rho(t)$ implies that $f(t)$ should be a continuous function, and also that $|f(t)|=1$. Since $\sum_{k}\hat{A}_{k}^{\dagger}(t)\hat{A}_{k}(t)=\hat{\mathcal{I}}$ for the map to be trace-preserving for any initial state $\hat{\rho}(0)$, it must hold that $0\le |f(t)|\le 1$. For instance, if  $\hat{\rho}(0)=|e\rangle\langle e|$, then
\begin{equation}
	\hat{\rho}(t)=\left(1-|f(t)|^{2}\right)|g\rangle\langle g|+|f(t)|^{2}|e\rangle\langle e|.
	\label{EqTimeEvol2LevelGeneric}
\end{equation}
More generally, it holds that
\begin{eqnarray}
\hat\rho(t) &=& |f(t)|^2\rho_{ee}(0)|e\rangle\langle e|+[\rho_{g}(0)+(1-|f(t)|^2)\rho_{ee}]|g\rangle\langle g|\nonumber\\
&& + f(t)\rho_{eg}(0)|e\rangle\langle g|+f^*(t)\rho_{ge}(0)|g\rangle\langle e|.
\end{eqnarray}
If we for simplicity assume that $f(t)$ is real, it is possible to show that this time evolution corresponds to a time-local Lindblad-like master equation~\cite{AnderssonCresserHall}
\begin{equation}
	\dot{\hat{\rho}}(t)=-\frac{\dot{f}(t)}{f(t)}\left[2\hat{\sigma}_{+}\hat{\rho}(t)\hat{\sigma}_{-}-\hat{\sigma}_{+}\hat{\sigma}_{-}\hat{\rho}(t)-\hat{\rho}(t)\hat{\sigma}_{+}\hat{\sigma}_{-}\right],
	\label{EqMaster2LevelAtom}
\end{equation}
where $\hat{\sigma}_{+}=|e\rangle\langle g|$ and $\hat{\sigma}_{-}=|g\rangle\langle e|$. (For complex $f(t)$, the master equation contains an additional term which is unimportant for our purposes.) We can thus identify the generally time-dependent decay rate, for transitions from $|e\rangle$ to $|g\rangle$, as $\gamma(t)=-2\dot{f}(t)/f(t)$. The choice $f(t)=\exp(-\gamma t)$, where $\gamma>0$ is a constant, corresponds to the familiar case of exponential decay with a decay rate $\gamma$. An $|f(t)|$ which is monotonically decreasing in some other way corresponds to a decay rate that varies as a function of time. If $|f(t)|$ is increasing, then this corresponds to a negative decay rate and non-Markovian behaviour.

The time evolution becomes non-invertible if $f(t)=0$ for certain time(s)~\cite{AnderssonCresserHall}. Any initial state will evolve to $|g\rangle\langle g|$ at these times. One such choice is  $f(t)=\cos(\omega t)$, which as we shall see corresponds to the Jaynes-Cummings model on resonance, describing the interaction of an atom with a cavity field. We then have $\gamma(t)=2\omega\tan(\omega t)$, which not only becomes negative, but also diverges when the time evolution becomes non-invertible, that is, for $\omega t=(n+1/2)\pi$ where $n\in \mathbb{Z}$.  One easily sees that the solution of the corresponding master equation is not unique. Another function $f(t)=A \cos(\omega t)$, where  $0<|A|<1$ and $A$ is a constant, would also give $\gamma(t)=2\omega\tan(\omega t)$ and exactly the same master equation (\ref{EqMaster2LevelAtom}), but a different time evolution. Conversely, if we know that the initial state of the system is something other than $|g\rangle\langle g|$, then it is possible to fix the value of $|A|$. However, if the initial state is $|g\rangle\langle g|$, or if this state is encountered during the time evolution, then we cannot determine $|A|$ for the subsequent time evolution. The system could start its time evolution with one value of $A$, and continue evolving with another value of $A$ after having reached the ground state. Such a time evolution would be a solution to the time-local master equation in \eqref{EqMaster2LevelAtom}, just as valid as a solution with an $A$ that remains constant. 

We will proceed to explicitly construct Hamiltonians corresponding to two time evolutions that correspond to the same time-local master equation for this example.

\subsection{Hamiltonian and time evolution}
\label{sec:hamiltonian_and_time_evolution}

The Jaynes-Cummings Hamiltonian for a two-level atom coupled to a quantised electromagnetic field is given by
\begin{equation}
	\hat{\mathcal{H}}_{\text{JC}}=\hbar\omega\left(\hat{a}^{\dagger}\hat{a}+\frac{1}{2}\right)+\frac{1}{2}\hbar\omega_{0}\hat{\sigma_{z}}-i\hbar\Omega\left(\hat{\sigma}_{+}+\hat{\sigma}_{-}\right)\left(\hat{a} - \hat{a}^{\dagger}\right),
	\label{EqJCHam}
\end{equation}
where $\hat a^\dagger , \hat a$ are the creation and annihilation operators for the field, $\omega$ is the frequency of the field, $\hbar\omega_0$ is the energy difference between $|g\rangle$ and $|e\rangle$, and $\Omega$ the strength of the coupling between the two-level system and the cavity field. In the interaction picture, and after making the rotating wave approximation (RWA), the Jaynes-Cummings Hamiltonian becomes~\cite{JaynesCummings}
\begin{equation}
	\hat{\mathcal{H}}_{\text{int}}=\hbar\Delta\hat{\sigma}_{z}-i\hbar\Omega\left(\hat{\sigma}_{+}\hat{a}-\hat{\sigma}_{-}\hat{a}^{\dagger}\right),
	\label{EqJHInt}
\end{equation}
where $2|\Delta|=\omega_0-\omega$ is the detuning of the system. The RWA is a good approximation if $2\Delta\ll \omega_0+\omega$ and $\Omega\ll \omega_0$.

In order to use dimensionless units, we scale variables by the time $\tau$ as
\begin{eqnarray}
	\tilde{t} & = & \frac{t}{\tau}, \\
	\widetilde{\omega}_{0} & = & \tau \omega_{0}, \\
	\widetilde{\omega} & = & \tau \omega, \\
	\widetilde{\Omega} & = & \tau \Omega, \\
	\widetilde{\Delta} & = & \tau \Delta.
\end{eqnarray}
The time evolution is determined by the time-dependent Schr\"odinger equation
\begin{equation}
	i\hbar\frac{\partial}{\partial \tilde{t}}|\Psi(\tilde{t})\rangle=	\hat{\mathcal{H}}_{\text{int}}|\Psi(\tilde{t})\rangle.
	\label{EqSchroedingerEq}
\end{equation}
In the one-excitation subspace, the basis states for the atom and the field are $|g,1\rangle$ and $|e,0\rangle$. Within this subspace, the time-dependent state can be expressed as
\begin{equation}
	|\Psi(\tilde{t})\rangle=c_{e}(\tilde{t})|e,0\rangle+c_{g}(\tilde{t})|g,1\rangle,
	\label{EqJointState}
\end{equation}
where $c_{e(g)}(\tilde{t})$ are time dependent coefficients. The Schr\"odinger equation expressed in matrix form is
\begin{equation}
	i\tau\frac{\partial}{\partial \tilde{t}}{{c_e(\tilde{t})}\choose{c_g(\tilde{t})}}=\left(\begin{array}{ccc}\widetilde{\Delta} & -i\widetilde{\Omega} \\ i \widetilde{\Omega} & -\widetilde{\Delta} \end{array}\right){{c_e(\tilde{t})}\choose{c_g(\tilde{t})}}.
	\label{EqHamiltonianMatrix}
\end{equation}
For instance, for $c_{g}(0)=1$ and $c_{e}(0)=0$ the solution is
\begin{eqnarray}
	\label{EqJointStateSolutionInitCond}
	|\Psi(\tilde{t})\rangle&=&-\frac{\widetilde{\Omega}}{\widetilde{\omega}_{R}}\sin\left(\widetilde{\omega}_{R} \tilde{t}\right)|e,0\rangle  \\
	&&+\left[\cos\left(\widetilde{\omega}_{R}\tilde{t}\right)+i \frac{\widetilde{\Delta}}{\widetilde{\omega}_{R}}\sin\left(\widetilde{\omega}_{R} \tilde{t}\right)\right]|g,1\rangle,\nonumber
\end{eqnarray}
where $\widetilde{\omega}_{R}=\sqrt{\widetilde{\Delta}^{2}+\widetilde{\Omega}^{2}}$ is the scaled Rabi frequency.

By varying $\widetilde{\Delta}$ and $\widetilde{\Omega}$ but keeping the Rabi frequency $\widetilde{\omega}_{R}=\sqrt{\widetilde{\Delta}^{2}+\widetilde{\Omega}^{2}}$ constant, we will now construct two different Jaynes-Cummings Hamiltonians that lead to different time evolutions but arbitrarily similar master equations. Evolution on resonance, with $\widetilde{\Delta}=0$, gives $\widetilde{\omega}_R=\widetilde{\Omega}$. An initial state $|e,0\rangle$ then gives $\hat{\rho}_{ee}(\tilde{t})=|c_e(\tilde{t})|^2=\cos^{2}(\widetilde{\omega}_R \tilde{t}), \hat{\rho}_{gg}(\tilde{t})=|c_g(\tilde{t})|^2=\sin^{2}(\widetilde{\omega}_R \tilde{t})$ and $\rho_{eg}=\rho_{ge}=0$ for the reduced density matrix of the atom. This will correspond to the first one of the two time evolutions (and Hamiltonians) in our example. The second time evolution starts off from the same initial state, also on resonance. At a time $\tilde{t}_i=(n+1/2)\pi/\widetilde{\omega}_R$, when the state is $|g,1\rangle$, the Hamiltonian is rapidly switched to off resonance with $\widetilde{\Delta}^\prime\neq 0$, also adjusting $\widetilde{\Omega}^\prime$ in such a way that the Rabi frequency $\widetilde{\omega}_R$ remains equal to $\widetilde{\Omega}$. If the switch is instantaneous, the reduced density matrix for the atom will after this time evolve with
\begin{eqnarray}
\rho_{ee}(\tilde{t})&=&|c_e(\tilde{t})|^2=\left|\frac{\widetilde{\Omega}'}{\widetilde{\omega}_R}\right|^2\cos^2(\widetilde{\omega}_R \tilde{t}),\nonumber\\
\rho_{gg}(\tilde{t})&=&|c_g(\tilde{t})|^2=1-\left|\frac{\widetilde{\Omega}'}{\widetilde{\omega}_R}\right|^2\cos^2(\widetilde{\omega}_R \tilde{t}),\nonumber\\
\rho_{eg}&=&\rho_{ge}=0.
\label{EqRhoTheory}
\end{eqnarray}
If the change is instantaneous and occurs at the right time, then the master equations for these two situations will be identical. However, if the change is not instantaneous, it will affect the master equation. We now proceed to numerically investigate how rapid changes to the Hamiltonian affect the time evolution of the atom and the corresponding master equation.

\subsection{Evolution with time-dependent coupling}
	\label{ssub:time_dependent_coupling}

\begin{figure}
    \centering{}
    \includegraphics[scale=0.65]{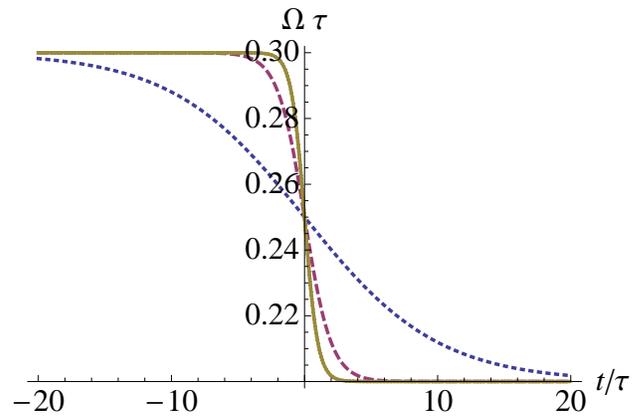}\\
    \caption{\label{FigCoupling}(Color online) Scaled coupling strengths $\widetilde\Omega =\Omega\tau$ for different rates of change, varying according to Eq. (\ref{EqCoupling}), with $\widetilde t_i =t_i/\tau=0$. Here, $\widetilde\Omega_{\rm max}=\tau\Omega_{\text{max}}$ is 0.3 and $\widetilde\Omega_{\rm min}=\tau\Omega_{\text{min}}$ is 0.2. The scaled values of $\widetilde{k}=\tau k$ are 0.1 for the dotted line (blue), 0.5 for the dashed line (red) and 1 for the solid line (gold). 
    }
\end{figure}

In order to change from an on-resonance to an off-resonance Hamiltonian with fixed $\omega_{R}$, both the coupling $\Omega$ and the detuning $\Delta$ must change. Let $\Omega(t)$ go from a maximum value $\Omega_{\text{max}}$ to a minimum $\Omega_{\text{min}}$. We choose a smooth step function to achieve this,
\begin{equation}
	\Omega(t)=\frac{\Omega_{\text{max}}-\Omega_{\text{min}}}{2}\left\{1-\tanh\left[k\left(t-t_i\right)\right]\right\}+\Omega_{\text{min}},
	\label{EqCoupling}
\end{equation}
where $k=\widetilde{k}/\tau$ controls the rate of change and $t_{i}$ is the point at which $\Omega(t)$ takes its average value $(\Omega_{\text{max}}+\Omega_{\text{min}})/2$, see Fig.~\ref{FigCoupling}. The detuning changes as $\Delta(t)=\sqrt{\omega_{R}^{2}-\Omega^{2}(t)}$, with $\omega_{R}$ kept constant. Changing the coupling strength may be feasible in cavity QED~\cite{RaimondHaroche}, by changing the position of an atom in a laser cavity, or by changing the laser field strength. In circuit QED, it is possible to realize tunable resonators~\cite{Palacios,Yamamoto}.

\begin{figure}
    \centering
    \includegraphics[scale=0.65]{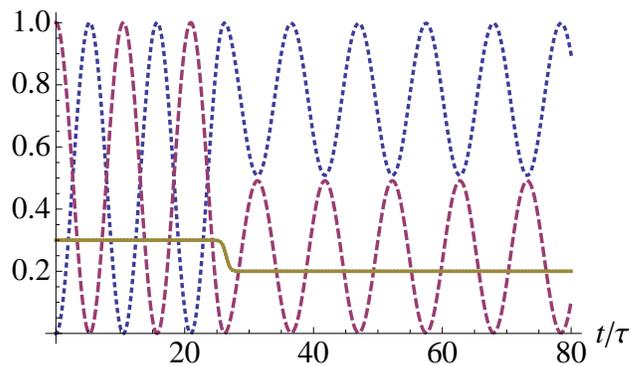}\\
    \caption{\label{FigNumSolution}{(Color online) Time evolution of the reduced density matrix elements of the atom for changing detuning $\widetilde\Delta(t)$ and coupling $\widetilde\Omega(t)$, with constant $\widetilde{\omega}_R$. The dotted line} (blue) shows $\rho_{gg}(\tilde{t})$ and the dashed line (red) $\rho_{ee}(\tilde{t})$. The parameter controlling the rate of change of the coupling $\Omega(t)$ is given by $\widetilde{k}=1.6$. The solid line (gold) shows how the coupling is changing.}
\end{figure}

\begin{figure}
	\centering
	\includegraphics[scale=0.65]{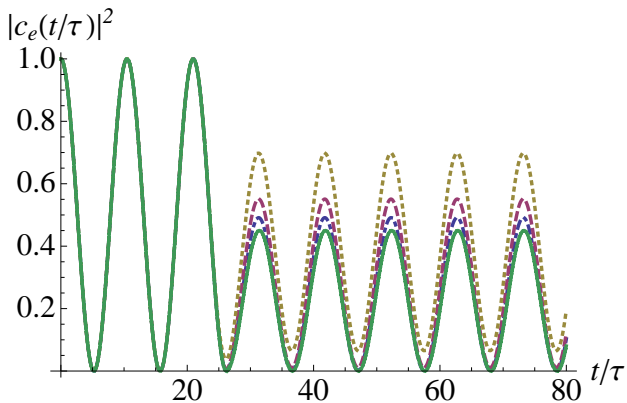}
	\caption{\label{PlotTimeEvK}(Color online) Time evolution of $\rho_{ee}(t)$ for different values of $\widetilde{k}$. The dotted curve (gold) corresponds to $\widetilde{k}=0.5$, in which case the system never reaches the ground state. As will be seen later, this affects the decay rate. The dashed curve (red) corresponds to $\widetilde{k}=1.0$ and the dot-dashed (blue) one to $\widetilde{k}=1.6$. The solid line (green) corresponds to $\widetilde{k}\rightarrow\infty$, that is, an instant change.}
\end{figure}

The numerical solution of (\ref{EqHamiltonianMatrix}) with $\Omega(t)$ given in \eqref{EqCoupling}, where $\widetilde{k}=1.6$, $\tilde t_i$ chosen so that $c_g(\tilde{t}_i)=0$, $\widetilde{\Omega}_{\text{max}}=0.3$, $\widetilde{\Omega}_{\text{min}}=0.2$, and $c_{g}(0)=0$ and $c_{e}(0)=1$, is plotted in Fig.~\ref{FigNumSolution}. It is seen that this produces the desired time evolution.
If the change in the coupling is not instantaneous, then this can be seen in the resulting time evolution.  Fig.~\ref{PlotTimeEvK} shows $\rho_{ee}(t)$ for different values for $k$. As the change becomes more rapid, the resulting dynamics approach those for an instantaneous change, see Fig.~\ref{PlotTimeEvK}. As $k$ decreases, and the change becomes faster, the amplitude of the oscillations after the change in the Hamiltonian decreases. 

\begin{figure}
    \centering
    \includegraphics[scale=0.65]{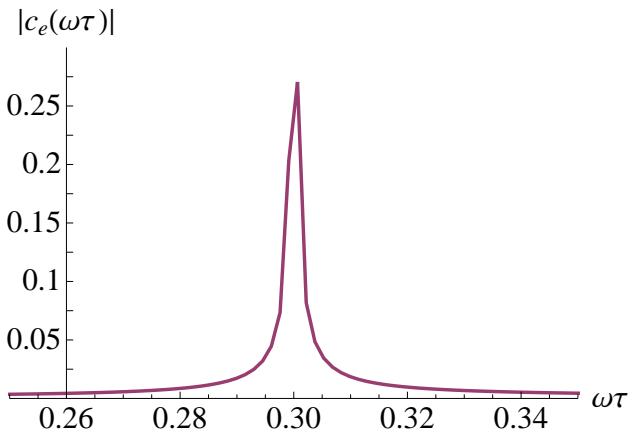}\\
    \caption{\label{FigPowerSpectrum} (Color online). Fourier analysis of $c_e(t)$, with $\widetilde{k}=1.6$, $\widetilde{\Omega}_{\text{max}}=0.3$ and $\widetilde{\Omega}_{\text{min}}=0.2$. When $k\rightarrow\infty$, $c_{e}(t)$ and $c_{g}(t)$ are piecewise defined functions, with  $c_{e}(t)=\cos(\omega_R t)$ if $t<t_i$ and $c_{e}(t)=A\cos(\omega_R t)$ if $t>t_i$, with $0<A<1$, and $c_{g}(t)=\sqrt{1-|c_{e}(t)|^2}$ (up to a phase sign). Fourier analysis of $c_{e}(t)$ when $k\rightarrow \infty$ gives a delta-function peak at $\widetilde{\omega}_R=\tau\omega_R=0.3$, together with a term proportional to $1/(\tau[\omega-\omega_R])$.}
\end{figure}

Fourier analysis of $c_e(t)$ and $c_g(t)$, see Fig.~\ref{FigPowerSpectrum}, confirms that the frequency of the oscillations in $\rho_{ee}$ stays constant to a good approximation. We have also numerically confirmed that the decay rate in the corresponding time-local master equation is practically indistinguishable from the decay rate for the case when the Hamiltonian is not changing in time. We therefore have an explicit example where two different time evolutions correspond to the same time-local master equation.
The procedure is nevertheless somewhat artificial in that it involves a rapid, ideally instantaneous, change in one of the Hamiltonians at a very specific time in order to get the desired time evolution and an unchanged decay rate. If the coupling and detuning do not change at the right time or rapidly enough, then $\rho_{ee}$ does not reach zero exactly, and then the time evolution will be invertible. Even if $\rho_{ee}$ does reach zero, if the change in the Hamiltonian is not instantaneous but occurs over a short but finite time period, then the master equation will not remain identical to the case where the Hamiltonian does not change, remaining on resonance.

\subsection{How rapidly should the coupling change?}
\label{sub:change_of_the_coupling}
We now discuss a practical estimate of how quickly the coupling should change for the time evolution to be experimentally indistinguishable from the case in which a change is made simultaneously. If a Hamiltonian changes from $\hat{\mathcal{H}}(t_0)=\hat{\mathcal{H}}_0$ to $\hat{\mathcal{H}}(t_1)=\hat{\mathcal{H}}_1$ between the times $t_0$ and $t_1$, then the probability that the state of the system remains unchanged from its initial state $|\psi_0\rangle$, is given by $1-\xi$, with
\begin{equation}
	\xi=\frac{T^2}{\hbar^2}\left(\langle \psi_0|\overline{\mathcal{H}}^2|\psi_0\rangle-\langle\psi_0|\overline{\mathcal{H}}|\psi_0\rangle^2\right),
\end{equation}
where $T=t_1-t_0$ and 
\begin{equation}
	\overline{\mathcal{H}}=\frac{1}{T}\int_{t_0}^{t_1}\hat{\mathcal{H}}(t)dt.
\end{equation}
If $\xi\ll 1$, then the system evolves diabatically, that is, the change in the Hamiltonian is so rapid that the system does not have time to adjust. 
It is straightforward to show that this leads to $\widetilde{k}\gg(\widetilde{\Omega}_{\text{max}}+ \widetilde{\Omega}_{\text{min}})/2$. Experimentally, it may be challenging to implement a rapid and precise enough change in the coupling strength and detuning.


\subsection{Decay rate in the time-local master equation}
\label{sec:decay_rate}

\begin{figure}
    \centering
    \includegraphics[scale=0.65]{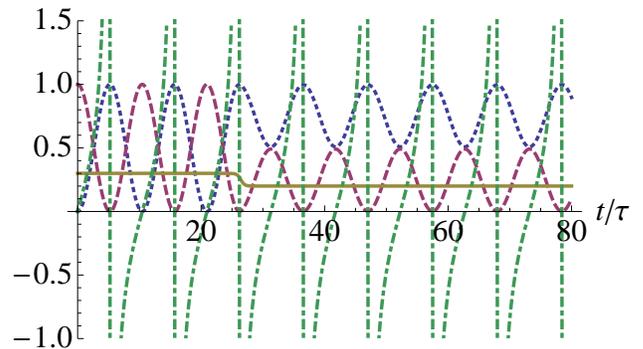}
    \caption{\label{FigPlotGeneral}(Color online) This figure shows the numerically obtained decay rate, the time evolution of the reduced density matrix elements and the change in the coupling for our example with the Jaynes-Cummings model. The dotted line (blue) and dashed line (red) correspond to $\rho_{gg}(t)$ and $\rho_{ee}(t)$ respectively. The dot-dashed line (green) represents the tangent-like decay rate and the solid line (gold) the time-dependent coupling with $\widetilde{\Omega}_{\text{max}}=0.3$, $\widetilde{\Omega}_{\text{min}}=0.2$ and $\widetilde{k}=1.6$.}
\end{figure}

\begin{figure}
	\centering
	\includegraphics[scale=0.65]{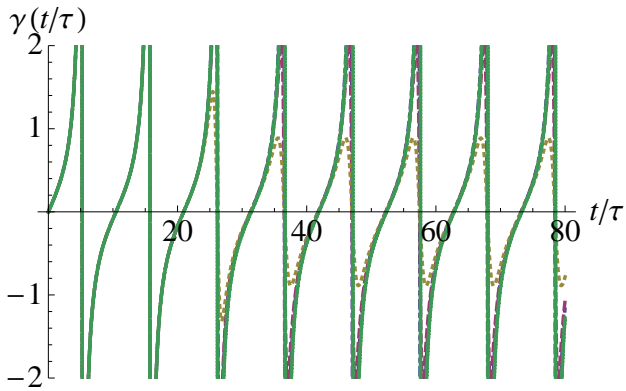}
	\caption{\label{FigDecayRateVarios}(Color online) Decay rates for different values of $\widetilde{k}$. The dotted curve (gold) corresponds to $\widetilde{k}=0.5$, the dashed curve (red) corresponds to $\widetilde{k}=1.0$ and the dot-dashed line (blue) to $\widetilde{k}=1.6$. The solid line (green) corresponds to $\widetilde{k}\rightarrow\infty$.}
\end{figure}

From the numerical solution of the Schr\"odinger equation for the case where the coupling changes, we can also compute the decay rate in the corresponding time-local master equation. The result is shown in Figs.~\ref{FigPlotGeneral} and~\ref{FigDecayRateVarios}.
As expected, the decay rate as a function of time approaches the functional form expected for an instantaneous change $\gamma(t)=2\omega_R\tan{\omega_{R}t}$, as the change becomes more rapid. This means that by looking only at the decay rate, we cannot distinguish between a Hamiltonian that does not change in time, and a Hamiltonian with a rapid enough change at the right time. From Fig.~\ref{FigPlotGeneral}, we also see that the decay rate indeed diverges when the time evolution is non-invertible, i.e. when the system is in the ground level. Also, a negative decay rate corresponds to ``re-coherence'' of the system and non-Markovian evolution.

\section{Conclusions} 
\label{sec:conclusions}
We have investigated time-local master equations through an example involving a two-level system. We constructed two different system-environment Hamiltonians, corresponding to two different time evolutions for the system, which nevertheless both give the same time-local master equation. This explicitly shows that the time-local master equation on its own is not enough to solve for the time evolution, if the time evolution is not invertible.

The example is nevertheless somewhat artificial since it involves rapid changes in one of the system-environment Hamiltonians at a specific instant in time. If any Hamiltonian is guaranteed to be ``physically well-behaved", for example, that it is continuous and does not change too fast (or more precisely, that its matrix elements are Lipshitz-continuous), one may conjecture that the corresponding time-local master equation does determine the time evolution, at least in principle, even when the time evolution is not invertible. This would in other words mean that two ``physically well-behaved" system-environment Hamiltonians, corresponding to different time evolutions for a quantum system when its environment is traced out, cannot both lead to the same master equation for the system. Equivalently, this would mean that if a master equation does not have a unique solution, then at least one of the solutions corresponds to an ``ill-behaved" system-environment Hamiltonian involving rapid changes. 

Now, as our example shows, even a ``well-behaved" Hamiltonian, such as the one in the Jaynes-Cummings model on resonance, may result in divergencies in the decay rate in the corresponding master equation. The usual theorems related to the existence and uniqueness of solutions of differential equations are of little help in proving our conjecture. The Picard-Lindel\"of theorem, for example, just tells us that the solution of such a master equation, with diverging decay rates, is not unique. If the decay rates in a Lindblad-like master equation do not diverge, then its solution would be unique -- but this simply corresponds to the case where the time evolution is always invertible. When the time evolution is not invertible, then decay rates will inevitably diverge at these times, and as already stated, this can and does happen even for very well-behaved system-environment Hamiltonians.	

If the time evolution is not invertible, then even if our conjecture holds true, and if the Hamiltonian is well-behaved enough for the time evolution to be uniquely defined by the time-local in a formal sense, one would still need to take care when numerically solving a time-local master equation. This is because the diverging decay rates may lead to instabilities in numerical calculations. Nonetheless, our results would generally support the view that time local master equations are applicable to a wider class of problems than one might expect on first inspection.



%

\end{document}